\documentclass[aps, prl, twocolumn, reprint,preprintnumbers, nofootinbib, showpacs, superscriptaddress,letterpaper]{revtex4}
\usepackage{amsmath}
\usepackage{amsfonts}
\usepackage{amssymb,multirow,wrapfig}
\usepackage{color}
\usepackage{graphicx}
\usepackage{cancel}
\usepackage{adjustbox}
\usepackage{array}
\usepackage{booktabs}
\usepackage{multirow}
\usepackage{lineno}

\newcommand{\ket}[1]{\ensuremath{\left| #1 \right>}}

\newcommand{\be}[0]{\begin{equation}}
\newcommand{\ee}[0]{\end{equation}}
\newcommand{\bea}[0]{\begin{eqnarray}}
\newcommand{\eea}[0]{\end{eqnarray}}

\begin{document}

\title{A quantum router for high-dimensional entanglement}

\author{Manuel Erhard}
\affiliation{Institute for Quantum Optics and Quantum Information (IQOQI), Austrian Academy of Sciences, Boltzmanngasse 3, A-1090 Vienna, Austria}
\affiliation{Faculty of Physics, University of Vienna, Boltzmanngasse 5, 1090 Vienna, Austria}

\author{Mehul Malik}
\email{mehul.malik@univie.ac.at}
\affiliation{Institute for Quantum Optics and Quantum Information (IQOQI), Austrian Academy of Sciences, Boltzmanngasse 3, A-1090 Vienna, Austria}
\affiliation{Faculty of Physics, University of Vienna, Boltzmanngasse 5, 1090 Vienna, Austria}

\author{Anton Zeilinger}
\affiliation{Institute for Quantum Optics and Quantum Information (IQOQI), Austrian Academy of Sciences, Boltzmanngasse 3, A-1090 Vienna, Austria}
\affiliation{Faculty of Physics, University of Vienna, Boltzmanngasse 5, 1090 Vienna, Austria}

\date{\today}

\begin{abstract} 

In addition to being a workhorse for modern quantum technologies, entanglement plays a key role in fundamental tests of quantum mechanics. The entanglement of photons in multiple levels, or dimensions, explores the limits of how large an entangled state can be, while also greatly expanding its applications in quantum information. Here we show how a high-dimensional quantum state of two photons entangled in their orbital angular momentum can be split into two entangled states with a smaller dimensionality structure. Our work demonstrates that entanglement is a quantum property that can be subdivided into spatially separated parts. In addition, our technique has vast potential applications in quantum as well as classical communication systems.


\end{abstract}

\maketitle

Considered one of the most seminal experiments in quantum physics, the Stern-Gerlach experiment provided the first direct evidence of quantization in quantum mechanics \cite{SG}. It is now well known that in addition to their spin, photons can also carry quantized values of orbital angular momentum (OAM) \cite{Allen:1992by}. Unlike spin, the OAM of a photon can take on a large range of discrete values. In direct analogy to the Stern-Gerlach experiment with atoms, recent demonstrations have shown that single photons carrying OAM can be spatially separated into directions corresponding to discrete linear momenta \cite{Leach:2002wy,Berkhout:2010cb,Mirhosseini:2013em}. The OAM of light also enabled the first laboratory studies of high-dimensional entanglement \cite{Mair:2001ub}, with recent experiments demonstrating entangled states of ever-increasing dimensionality \cite{Dada:2011vc,Krenn:2014jy, Malik:2016gu}. In this letter, we show how a quantum state composed of many entangled modes can be split into two smaller entangled states with exactly opposite parity. The original Stern-Gerlach experiment demonstrated that a single particle has discrete quantum properties that can be physically separated. Similarly, we show that entanglement, which is a property unique to quantum systems of two or more particles, can also be separated into two distinct parts.\\
\indent The orbital angular momentum of light is a physical property that appears due to a spatially varying field distribution \cite{Allen:1992by}. Photons carrying OAM have a helical wavefront where the phase, given by $\text{exp}(i \ell \phi)$, winds azimuthally around the optical axis. The number of twists $\ell$ in a $2\pi$ period indicate the amount of OAM carried by one photon in units of $\hbar$ \cite{Yao:2011ve}. The OAM state space is discrete and unbounded, which makes it particularly suitable as an information carrier in both quantum and classical communication \cite{Malik:2014ht}. Photons carrying OAM have been used to push the capacity of a quantum key distribution system beyond one bit per photon \cite{Mirhosseini:2015fy} and for reaching dazzling data rates in classical communication systems \cite{Ren:2016hu}. Photons entangled in their OAM have enabled superdense quantum communication \cite{Barreiro:2008hg} and were recently used to teleport a photon in a hybrid polarization-OAM space \cite{Wang:2015dm}.\\
\begin{figure}[b!]
\centering
\includegraphics[width=.9\linewidth]{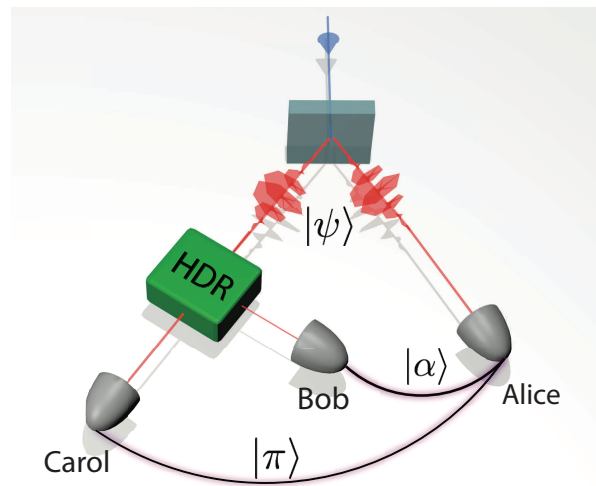}
\caption{A high-dimensional two-photon entangled state $\ket{\psi}$ is split into two smaller entangled states, $\ket{\pi}$ and $\ket{\alpha}$, and routed to three parties by a device called a high-dimensional entanglement router (HDR).}
\label{fig:hd-router-general}
\end{figure}
\indent Here we demonstrate how a high-dimensional OAM-entangled state $\ket{\psi}$ can be split into two smaller entangled states $\ket{\pi}$ and $\ket{\alpha}$ of different parity through the use of a high-dimensional entanglement router (Fig.~\ref{fig:hd-router-general}). This device uses an interferometric sorting method to distribute a high-dimensional entangled state originally shared between two spatially separated parties to three such parties. We quantify the entanglement dimensionality of these states and demonstrate that the high-dimensional non-classical correlations do indeed survive the process of subdivision. Furthermore, we demonstrate precise and stable control over the routing process for an extended period of time and show how the two entangled states can be exchanged between their respective parties on-demand and extremely fast. Besides being of use in fundamental tests, the entanglement router has immense potential applications in high-dimensional quantum communication systems \cite{Mirhosseini:2015fy,Krenn:2015ko}. Furthermore, this device can be readily used as an OAM switch in classical communication systems that exploit the spatial modes of light \cite{Ren:2016hu,Krenn:2014bx,Xie:2015jg}.
\indent The design of the high-dimensional entanglement router is based on an interferometer that was initially developed to sort single photons carrying OAM \cite{Leach:2002wy}. This device consists of a modified Mach-Zehnder interferometer with a dove prism in each arm, rotated at $90^\circ$ with respect to each other. The effect of a dove prism oriented at angle $\alpha$ is to rotate an incoming photon by an angle $2\alpha$. This introduces an OAM-dependent phase on the photon such that $|\ell\rangle\rightarrow \text{exp}(\pm 2 i \ell \alpha) |-\ell\rangle$ (note that a single reflection through the prism flips the OAM sign). Thus, a dove prism oriented at $90^\circ$ adds a phase of $\pi$ to photons carrying odd OAM and a phase of 0 to ones with even OAM (modulo $2\pi$). Sending an OAM-carrying photon through such an interferometer results in destructive or constructive interference (out of one port), depending on whether the photon was carrying odd or even values of $\ell$. In this manner, the interferometer sorts input photons based on the parity of their OAM quantum number. By changing the relative path length of the interferometer by $\pi$, the sorting direction can be changed.\\
\indent Consider an OAM-entangled 11-dimensional two-photon state, initially shared between two parties, Alice and Bob:

\be \ket{\psi}_{AB} = \sum_{\ell=-5}^5 c(\ell)\ket{\ell}_A\ket{-\ell}_B.\ee

\noindent Here, \textit{c}($\ell$) are complex probability amplitudes describing the state. The effect of sending one photon from this entangled state through the high-dimensional router is as follows:

\bea \ket{\psi}_{AB}&\xrightarrow{\textrm{HDR}}& \sum_{\ell=-5}^5 c(\ell)\ket{\ell}_A\big[\ket{-\ell}_B^\textrm{e}+\ket{-\ell}_C^\textrm{o}\big]\nonumber\\
&=&\sum_{\ell=-2}^2 c(2\ell)\ket{2\ell}_A\ket{-2\ell}_B\nonumber\\&&\;\;+\sum_{\ell=-3}^2 c(2\ell+1)\ket{2\ell+1}_A\ket{-2\ell-1}_C\nonumber\\
&=&\ket{\alpha}_{AB}+\ket{\pi}_{AC}.
\eea

\begin{figure}[t!]
\centering
\includegraphics[width=0.87\linewidth]{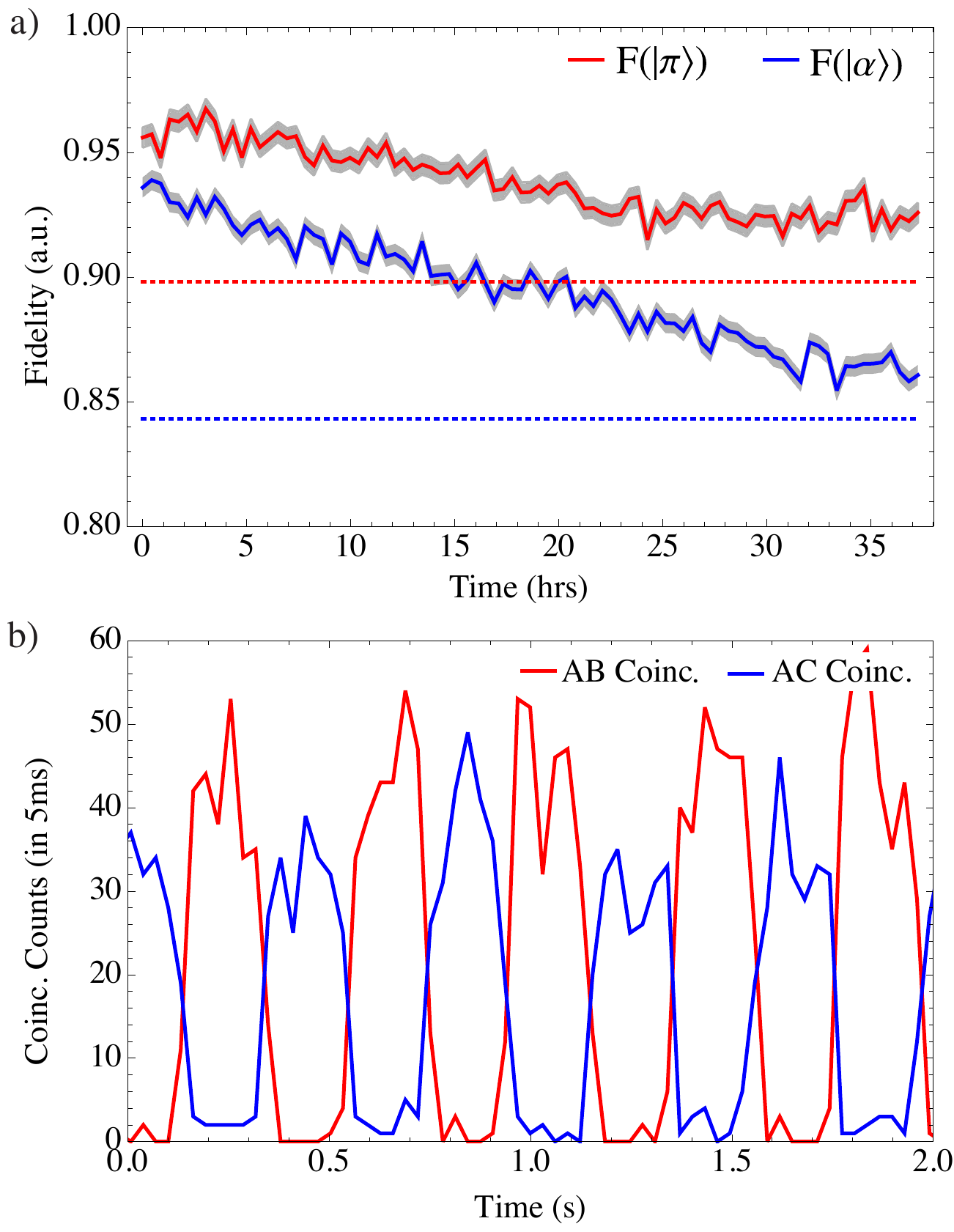}
	\caption{a) Long-term measurement of the experimental fidelity of $\ket{\pi}$ and $\ket{\alpha}$ demonstrating the presence of 6 and 5-dimensional entanglement respectively, over a period of 39 hours. The dotted lines represent the dimensionality bound. The error (shown in grey) is calculated via a Monte Carlo simulation of the experiment. b) Rapid switching of state $\ket{\alpha}$ between parties B and C at a frequency of 5Hz. Coincidence counts between detectors A+B (red) and A+C (blue) are displayed, with all detectors measuring an OAM superposition $|0\rangle+|2\rangle$.} 
\label{fig:longterm-data}
\end{figure}

\noindent The superscripts $e$ and $o$ represent even and odd modes respectively. Thus, the 11-dimensional state $\ket{\psi}_{AB}$ is split into a 5-dimensional entangled state $\ket{\alpha}_{AB}$ comprised of even OAM quanta $\ell\in\{-4,-2,0,2,4\}$, and a 6-dimensional entangled state $\ket{\pi}_{AC}$ comprised of odd OAM quanta $\ell\in\{-5,-3,-1,1,3,5\}$.\\
\begin{figure}[t!]
\centering
\includegraphics[width=\linewidth]{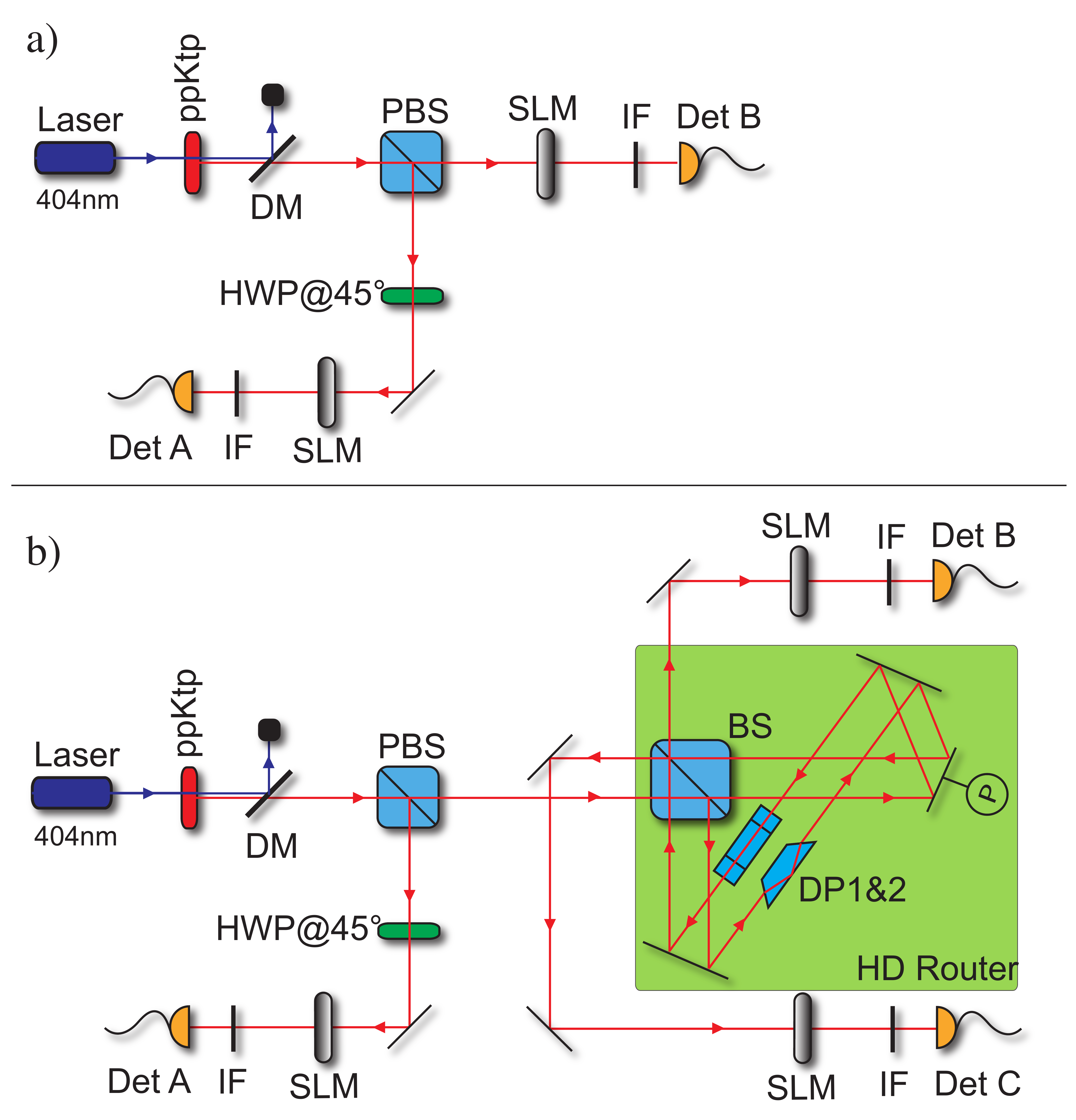}
	\caption{Schematic of experimental setups used for: a) generating and verifying a high-dimensional OAM-entangled state and, b) splitting the state into two entangled states with a smaller dimensionality and routing them to two different parties. Both setups use a pulsed laser source operating at 404nm to produce entangled photon pairs at 808nm in a nonlinear ppKTP crystal. Spatial light modulators (SLM) and single mode fibers are used in combination for detecting the OAM-entangled photonic state. Interference filters (IF) ensure a narrow bandwidth. The high-dimensional router (HD Router) is used to separate the odd and even OAM quanta of one of the photons, routing them to detectors B and C (details in the text).} 
\label{fig:hd-router-detailed}
\end{figure}
\begin{figure*}[t!]
\centering
\includegraphics[width=\textwidth]{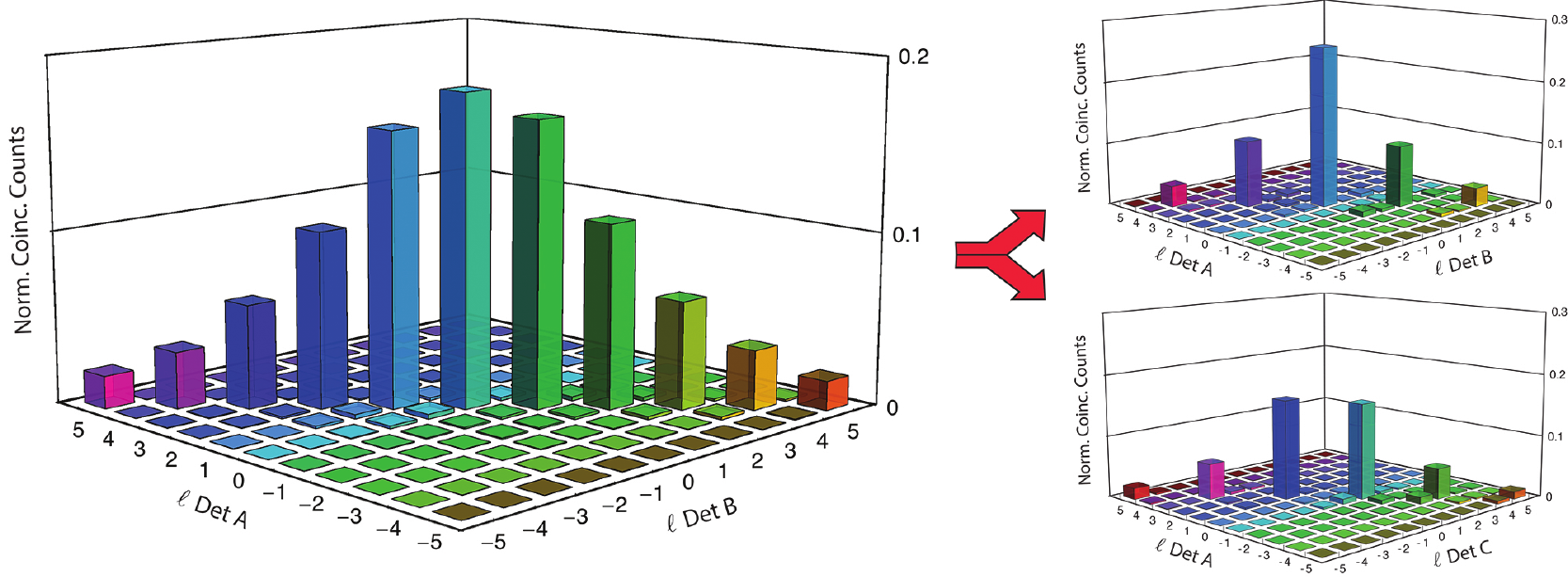}
	\caption{OAM correlations between the two photons measured before and after the routing process. The normalized coincidence counts between detectors A+B and A+C as a function of measured OAM mode are shown. As can be seen, the states are all strongly anti-correlated in OAM. However, the odd and even terms are deterministically separated after the routing process.} 
\label{fig:corr-data-all}
\end{figure*}
\indent First, we generate the high-dimensionally entangled state $\ket{\psi}_\textrm{AB}$ using the experimental setup depicted in Fig.~\ref{fig:hd-router-detailed}a). This consists of a femtosecond pulsed laser operating at 404nm that is used to pump a 1mm thick, periodically poled Potassium Titanyl Phosphate (ppKTP) crystal. The laser is focused into the crystal with a beam waist of 240$\mu$m. This ensures that the distribution of OAM values is very broad and that our state contains sufficient higher-order modes. Entangled photon pairs at 808nm are generated via a type-II collinear spontaneous parametric downconversion (SPDC) process, and are separated from the pump by a dichroic mirror (DM). A polarizing beam splitter (PBS) separates the two photons, directing each to a detection system consisting of a spatial light modulator (SLM), a single mode fiber (SMF) and a single photon detector. These devices are used in combination to perform projective measurements of OAM modes and their superpositions. The SLM flattens the phase of an incoming mode, allowing it to couple efficiently into the SMF. A home-built electronic circuit records coincidence counts between the two detectors. This allows us to make suitable measurements on the generated two-photon state and verify its entanglement dimensionality.\\
\indent The high-dimensional entanglement router (shown in the green square in Fig.~\ref{fig:hd-router-detailed}b) is inserted into arm B, effectively splitting it into two paths B and C. Instead of the original Mach-Zehnder design \cite{Leach:2002wy}, the HD router is implemented in a folded, double-path Sagnac configuration. The input beam is split into two counter-propagating Sagnac loops that meet back up the beam splitter. Since these two loops use the same mirrors and follow a similar path through the interferometer, path length fluctuations due to mechanical stresses or vibrations are effectively cancelled out. Each loop contains a dove prism (DP1 and 2). One of the mirrors of the interferometer is connected to a computer-controlled Piezo actuator (P) that allows fine control over the alignment. This feature is also used for rapid automatic realignment of the interferometer in order to compensate for temperature drifts. The piezo also allows us to rapidly switch the routing direction (\textit{even/odd}). The entire HD router is enclosed in a plastic box which reduces air flow and provides temperature stability to within $\pm0.02^\circ C$. This configuration provides a very high alignment stability and allows the router to operate over several days, as is shown later.\\
\indent The results of measurements performed in the OAM basis, both before and after routing, are shown in Fig.~\ref{fig:corr-data-all}. As can be seen in Fig.~\ref{fig:corr-data-all}a), the two-photon state is strongly anti-correlated in OAM. After passing through the HD router, the state is split into two smaller states containing only even (Fig.~\ref{fig:corr-data-all}b) and odd OAM terms (Fig.~\ref{fig:corr-data-all}c) that are still strongly anti-correlated in OAM. Note that these matrices only show all the diagonal elements of each state, which are insufficient to show entanglement. In order to prove that these states are indeed high-dimensionally entangled, we use an entanglement witness that puts a bound on the entanglement dimensionality of the state \cite{fickler2014interface}. First, we calculate the fidelity $F_{exp}$ between our state and an ideal target state $\ket{\phi_d}=\sum_{\ell=-L}^{L}c_\ell |-\ell,\ell\rangle$, which is $d$-dimensionally entangled ($d=2L+1$). Then, we calculate a $d$-dimensional entanglement bound $\mathcal{B}(d)=\sum_{i=0}^{d-1} \lambda_i$, which is the sum of all but the smallest Schmidt coefficient of the target state. If the calculated fidelity exceeds the bound $\mathcal{B}(d)$ for a $d$-dimensional entangled state, then the measured correlations can only be explained with a $d+1$ dimensional entangled state.\\ 
\begin{table}[b]
  \centering
  \caption{\label{tab:tab1} Measured fidelity $F_{exp}$ and estimated entanglement bounds $\mathcal{B}(d)$. 
  Errors are calculated using Monte Carlo simulation with Poissonian distribution of counting statistics.\vspace{1mm}}
  \begin{tabular*}{0.45\textwidth}{@{\extracolsep{\fill}} c|c|c|c}
  \hline\hline
 State & $F_{exp}$ & Bound $\mathcal{B}(d)$ & Dimensionality $d$ \\ \hline\hline
 $|\psi\rangle_{AB}$ & $0.757\pm0.002$ & $0.745$ & 10 \\\hline
 $|\alpha\rangle_{AB}$ & $0.937\pm0.004$ & $0.843$ & 5 \\\hline
 $|\pi\rangle_{AC}$  & $0.956\pm0.004$ & $0.898$ & 6 \\
 \hline\hline
 \end{tabular*}
\end{table}
\begin{figure*}[t]
\centering
\includegraphics[width=\textwidth]{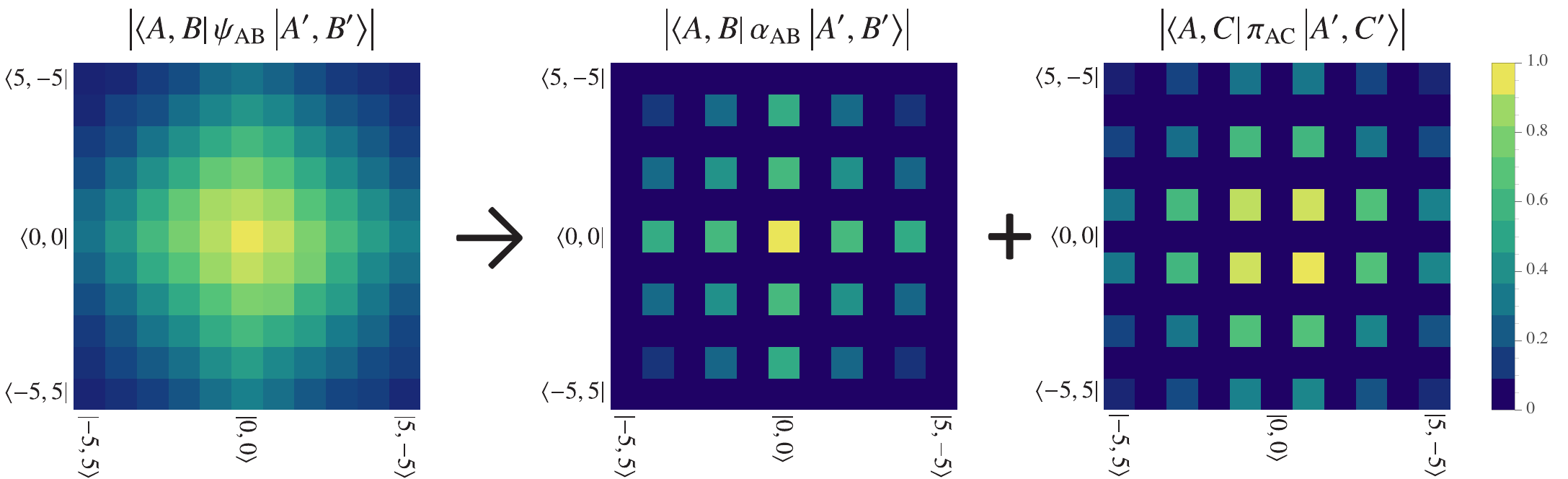}
	\caption{Measured density matrix elements (absolute values) showing how the entanglement router splits the state $\psi_\textrm{AB}$ into odd and even states $\alpha_\textrm{AB}$ and $\pi_\textrm{AC}$. Note that only the expected non-zero density matrix elements are necessary for calculating the fidelity and are measured. The elements of $\alpha_\textrm{AB}$ and $\pi_\textrm{AC}$ that are not measured are depicted in dark blue to illustrate the routing process.} 
\label{fig:DM}
\end{figure*}
\indent As is seen in Fig.~\ref{fig:corr-data-all}, the distribution of OAM modes is far from flat. Thus, it is advantageous to find a target state $|\phi_d\rangle$ which is close to the observed distribution. Note that the entanglement bound $\mathcal{B}(d)$ always leaves out the smallest Schmidt coefficient of $\ket{\phi_d}$, which is directly connected to its smallest probability amplitude. Thus, as the state gets more asymmetric, the entanglement bound gets tighter. In order to optimize this, we define a parameter $\xi=F_{exp}-\mathcal{B}(d)$ that measures the distance between the experimental fidelity $F_{exp}$ and the entanglement bound $\mathcal{B}(d)$. By maximizing $\xi$, we find the entanglement dimensionality $d$ of the measured state. Table 1 shows the measured fidelity and entanglement dimensionality of the states $\ket{\psi}_{AB}$, $\ket{\alpha}_{AB}$, and $\ket{\pi}_{AC}$. In order to calculate the fidelity of each state, it is sufficient to measure its density matrix elements that are expected to be non-zero. In the case of $\ket{\psi}_{AB}$, there are 121 such elements. These are plotted in Fig.~\ref{fig:DM}. In order to illustrate the sorting process, the density matrix plots of states $\ket{\alpha}_{AB}$ and $\ket{\pi}_{AC}$ also show the odd and even parity elements that are expected to be zero in dark blue. From these measurements, we are able to show that the state $\ket{\psi}_{AB}$ before the router has an entanglement dimensionality of 10. The state is then split by the HD router into two smaller states $\ket{\alpha}_{AB}$ and $\ket{\pi}_{AC}$, which are verified to have dimensionalities of 5 and 6 each. While we only show elements of $\ket{\psi}_{AB}$ up to $\ell=\pm5$, we needed to take the next higher order modes of $\ell=\pm6$ into account in order to achieve a dimensionality of 10. These results give rise to an interesting question: How can a 10-dimensionally entangled state be split up into a 5 and 6-dimensionally entangled state? The answer is that the state $\ket{\psi}_{AB}$ is very close to the bound of being 11-dimensionally entangled. The HD router acts like a filter for \textit{even/odd} modes, which results in a lower crosstalk between these modes after the routing process. This increases the visibility of the density matrix measurements and results in a higher quality state at both AB and AC.\\
\indent Finally, we perform long term measurements of the fidelity of states $\ket{\alpha}_{AB}$ and $\ket{\pi}_{AC}$ over a continuous period of 39 hours (Fig.~\ref{fig:longterm-data}a). During this time, both states have a fidelity that is well above the entanglement bound for their respective entanglement dimensionality. The HD router is calibrated every 26 minutes with the piezo crystal to ensure a high sorting visibility. The first fidelity measurements were taken to estimate the optimal target state $|\phi_d\rangle$, which was then kept the same for every successive measurement. Even though the HD router is continuously re-calibrated, the experimental fidelities decrease constantly. The total count rates are also seen to decrease steadily, which points to gray tracking of the ppKTP crystal as a possible source of error. Grey tracking would affect the mode quality of higher order OAM modes and would thus also lower the experimental fidelity over time. Fig.~\ref{fig:longterm-data}b demonstrates the switching capability of the HD router. By changing the relative path length difference of the interferometer with the piezo crystal, it is possible to choose whether Bob or Carol receive \textit{even} or \textit{odd} modes and thus the $|\alpha\rangle$ or the $|\pi\rangle$ state. The coincidence counts between AB and AC are shown with an OAM superposition of $|0\rangle+|2\rangle$ being measured at all three detectors. The piezo crystal is programmed such that it switches back and forth between two voltages ($V_1=31.5V, V_2=51.4V$) with a switching frequency of 5Hz, which is mainly limited by our count rates. The average visibility of the switching process is $0.97\pm 0.02$. In principle, one could achieve extremely fast switching frequencies by using a Pockels cell instead of a piezo crystal to introduce a phase shift.\\
\indent In summary, we have reported the development of a high-dimensional entanglement router that uses an interferometric method to split a high-dimensional entangled state into two smaller states, while maintaining the high-dimensional non-classical correlations that were present originally. This serves as an analog of the Stern-Gerlach effect when applied to the quantum property of entanglement. We use the entanglement router to divide a two-photon state entangled in 10 dimensions of its orbital angular momentum (OAM) into two states entangled in 5 and 6 dimensions of their OAM. Additionally, we demonstrate stable operation of this device over several days, and show how it can rapidly switch the routing direction. The entanglement router will find extensive application in quantum networks relying on high-dimensional entanglement, as well as in high-capacity classical communication links that exploit the OAM of light.

\noindent \textbf{Funding.} Austrian Science Fund (FWF), European Commission Marie Curie IIF (OAMGHZ), European Research Council (ERC).

\noindent \textbf{Acknowledgments.} We thank M. Krenn for fruitful discussions.

\end{document}